\documentclass[11pt]{article}

\usepackage{fullpage}
\usepackage[figuresright]{rotating}
\usepackage{amsmath}
\usepackage{times}
\usepackage{natbib}
\usepackage{graphicx} 
\usepackage{multirow}
\usepackage{arydshln}
\usepackage{hyperref}

\usepackage[shortlabels]{enumitem}

\newcommand{\blind}{1}

\def\bSig\mathbf{\Sigma}

\newcommand{\e}[1]{{E}\left[#1 \right]}

\begin{document}

\def\spacingset#1{\renewcommand{\baselinestretch}%
{#1}\small\normalsize} \spacingset{1}

\title{ \bf Designs with complex blocking structures and network effects for agricultural field experiments}

\if1\blind
{
\author
{\large Vasiliki Koutra$^{1}$\footnote{Contact: Vasiliki Koutra; \texttt{vasiliki.koutra@kcl.ac.uk}; Department of Mathematics,
King's College London, London WC2R 2LS, UK}, Steven G. Gilmour$^{1}$,
Ben M. Parker$^{2}$, and Andrew Mead $^{3}$\\[1ex]
$^{1}$Department of Mathematics, King's College London, London WC2R 2LS, UK \\
$^{2}$Department of Mathematics, Brunel University London, Uxbridge, UB8 3PH, UK \\
$^{3}$ Computational and Analytical Sciences, Rothamsted Research, Harpenden AL5 2JQ, UK}
} \fi

 \date{\vspace{-1.2cm}}

  \maketitle

\bigskip
\begin{abstract}
We propose a novel model-based approach for constructing optimal designs with complex blocking structures and network effects, for application in agricultural field experiments. The potential interference among treatments applied to different plots is described via a network structure, defined via the adjacency matrix. We consider a field trial run at Rothamsted Research and provide a comparison of optimal designs under various different models, including the commonly used designs in such situations. It is shown that when there is interference between treatments on neighbouring plots, due to the spatial arrangement of the plots, designs incorporating network effects are at least as, and often more efficient than, randomised row-column designs. The advantage of network designs is that we can construct the neighbour structure even for an irregular layout by means of a graph to address the particular characteristics of the experiment. The need for such designs arises when it is required to account for treatment-induced patterns of heterogeneity. Ignoring the network structure can lead to imprecise estimates of the treatment parameters and invalid conclusions. \\
\end{abstract}


\vfill

\section{Introduction}
\label{s:intro}

Agricultural field experiments often exhibit neighbour effects, that is, the responses to treatments on plots are affected by the treatments applied to neighbouring plots \citep{Cox1958}. For example, a chemical pesticide applied to one plot can potentially influence the responses on surrounding plots due to spray drift, a taller variety in one plot may influence the growth of a shorter variety on a neighbouring plot by shading the plants, the lack of control of a foliar disease on one plot may cause there to be a higher disease pressure on neighbouring plots, or the control or lack of control of an insect pest on one plot may influence the pest pressure on plants in neighbouring plots. Research on reducing or accounting for neighbour effects has mainly concentrated on the construction of either neighbour-balanced designs within crossed and nested blocking structures or designs that are optimal for some interference model (see, e.g. \citealp{DavidandKempton1996}). This work proposes the construction of optimal designs with complex blocking structures, while also accounting for neighbour effects. Although our focus is primarily on agricultural field experiments, the suggested methodology is applicable in a wide variety of medical, marketing and industrial contexts for obtaining efficient designs tailored to the particular experimental requirements and constraints (for experiments on networks, see, e.g. \citealp{Aral2016, Bapna2015, Bond2012, Centola2010}). For a marketing experiment on a social network, for example, we need to select which users should receive which advertisements of a commercial product in order to assess differences in click-through rates or revenue \citep{Xu2015}.

Blocking involves grouping together experimental units that are expected to have similar responses in the absence of treatments. Agricultural field experiments are often designed with simple (e.g. randomised complete block designs) or more complicated (incorporating nested and/or crossed structures) blocking structures to allow for anticipated systematic sources of variability associated with the physical arrangement of plots or with constraints on the management of groups of plots. Systematic sources of variability associated with the physical arrangement of plots might include trends in soil characteristics, such as pH or soil fertility, topography (e.g. locations down a slope), or distance from a field margin or hedge that might be a source for pests or diseases. Constraints imposed by farm management activities could include lines of plots being drilled in a single pass of the farm machinery, or a set of plots being simultaneously sprayed with nutrients or pesticides under a boom sprayer. Where there is only a single source of such variability, or where multiple sources can be confounded, then simple blocked designs can be used, but often these potential systematic sources of variability will affect different subsets of plots, so that more complex blocking structures need to be incorporated into the designs.

With relatively few treatments and the need to block for potential systematic variability in two orthogonal directions, standard row-column designs might be appropriate, such as Latin squares or Youden squares (where rows and/or columns contain a complete replicate of the treatment set).  But as the number of treatments increases, particularly where only relatively few complete replicates are possible (as is usually the case with agricultural field experiments), it will often be necessary to consider the impact of rows and columns of plots within each of a number of (replicate) blocks. In some cases, these blocks will be physically separated due to farm management constraints, whilst in other cases the blocks will be contiguous, additionally allowing for row and column structures to run across multiple blocks.  Thus, many agricultural field experiments have to be designed within large two-dimensional (row-column) arrays of plots, allowing for variation between rows and between columns. 

A commonly used approach are resolved row-column designs, where each block contains a complete replicate of the set of treatments \citep[Ch.4-6]{JohnandWilliams1995}. Several authors have developed resolvable row-column designs for comparing different treatments, for example \citet{Bose1947}, \citet{SinghandDey1979}, \citet{Ipinyomi1985}, \citet{Bailey1993} and \citet{Piecho2015}. In general, such designs are preferred when dealing with a large number of treatments (e.g.\ a large number of varieties) and a small number of replications. Other possibilities include $\alpha$-designs (introduced by \citealp{PattersonandWilliams1976}), which are resolvable block designs with respect to a single within-block sub-blocking component (see also \citealp{JohnandEccleston1986}, who developed a class of orthogonal row-column designs based on the $\alpha$-designs). The software package CycDesigN \citep*{Whitakeretal2002} is a practical tool for constructing efficient resolvable nested block and row-column designs. 

Much recent work on the design of experiments of agricultural experiments concerns spatial design for mixed models with complicated correlation structures (see, e.g. \citealp{Edmondson2020, Verdooren2020, Piecho2020}). \citet{Hoefler2020} described a large-scale simulation study to investigate the performance of spatial designs relative to designs with more traditional blocking structures across different scenarios. Their design evaluations revealed that replication improved the design performance and designs that controlled for some spatial variability had overall best performance. This literature focuses on the unit structure (including spatial effects). In the current paper, we instead account for treatment interference including spillover (neighbour) effects, additionally to a complicated blocking structure. 

Plot interference is another potential source of experimental error and bias which occurs when the response is affected by (either response or treatment) interference in neighbouring plots. Adjustments to the unit structure or modelling unit effects do not directly address the estimation of such interference effects, which are indirect treatment effects. There are practical methods for reducing the effects of plot interference, such as using wider spacing between plots, including border plants as guards, or using larger plots and only recording on inner rows \citep{Spitters1979, BesagandKempton1986}. 

Apart from standard experimental practices that aid in decreasing interference through implementation, it may be possible to take account of treatment interference effects by including additional terms in the response model and optimising the design for such a model. A wide variety of possible models and designs have been suggested for accommodating treatment interference. Examples include the work of \citet{DavidandKempton1996}, \citet{Druilhet1999}, \citet{KunertandMartin2000}, \citet{Bailey2004} and \citet{KunertandMersmann2011}, who developed models and efficient designs for experiments where units are arranged in a circle or a line allowing for the effects of immediate neighbours. Their suggested designs were primarily neighbour-balanced in the sense that all pairs of treatments occur in adjacent plots equally often in some cases allowing for the direction of any neighbour effects. Important work suggesting ways of accommodating interference in the analysis of field experiments is that of \citet{BesagandKempton1986}, which investigated different causes of association between neighbouring plots and provided appropriate models for better capturing each cause (i.e. spatial techniques for the adjustment of field variation, response interference--interplot competition, and treatment interference). 

More recently, \citet*{Parkeretal2017} suggested a model that relaxed the assumption of neighbour effects being controlled in only one direction and allowed for a network setting.  \citet*{Koutraetal2021} constructed efficient block designs using an extended version of that model with the inclusion of blocks in addition to the neighbour effects for eliminating heterogeneity across experimental units. Here, we extend the latter work to consider more complex blocking structures.

We obtain optimal resolved row-column designs that are suitable for use in agricultural field experiments when additionally there is an underlying interference structure governing the plots which can be captured via a network. That network structure is represented by means of a graph and accounts for sources of field variation from adjacent plots caused by he direct interference or competition effects due to the treatments, or indirect effects associated with crop management activities (e.g. drilling, harvesting), or by variation in allocated resources (e.g. variation in pesticide rates for plots treated simultaneously). The network structure therefore represents separate sources of variation from the systematic sources captured by conventional blocking structures. Thus, our suggested complex blocked designs with network effects aim at controlling heterogeneity from multiple sources ultimately maximising the separation of treatment information from within-field trends and other sources of variation which could influence the estimation of treatment effects and comparisons. 

Section \ref{sec:2} considers an agricultural field experiment that was  designed and implemented at Rothamsted Research and serves as a motivating example for this paper. The specification of the adjacency matrix for this example experiment is considered in this section. Section \ref{sec:4} provides the model on which the optimal designs will be based capturing the potential treatment interference between neighbouring units, incorporating both network effects and the nested and crossed blocking factors. The formulation of $A_s$-optimality for estimating differences between the treatment effects in the presence of nuisance network effects is also discussed in this section. Several optimal designs are provided in Section \ref{sec:6} for the motivating example field experiment with a detailed comparison made among them. Finally, Section \ref{sec:8} discusses relevant practical issues and concludes the paper.

\section{Motivating application}
\label{sec:2}

We consider an agricultural field experiment which aimed to assess differences in the natural cereal aphid colonisation of eighteen selected lines from the Watkins bread wheat landrace collection \citep{Wingen2014} compared to three elite wheat varieties. For practical reasons, the experiment was restricted to small plots (1m by 1m) for testing insect preference among the different wheat lines, with sufficient seed resources for 4 complete replicates, and space for an array of 84 plots arranged in 14 rows and 6 columns was available. The size of the plots and relatively small distances between neighbouring plots (0.75m between rows, 0.5m between columns) suggests the potential for direct interference effects because of lines having different levels of susceptibility to aphid infestation. 

\begin{figure}[h!]
\centering
\caption{Field layout and treatment allocation (numbers) for the motivating agricultural experiment at Rothamsted Research (year 2016). Blue/green shading indicates the superrows and light/dark shading indicates supercolumns considered in the development of alternative models and designs}\label{fig1}
    \includegraphics[width=0.7\textwidth]{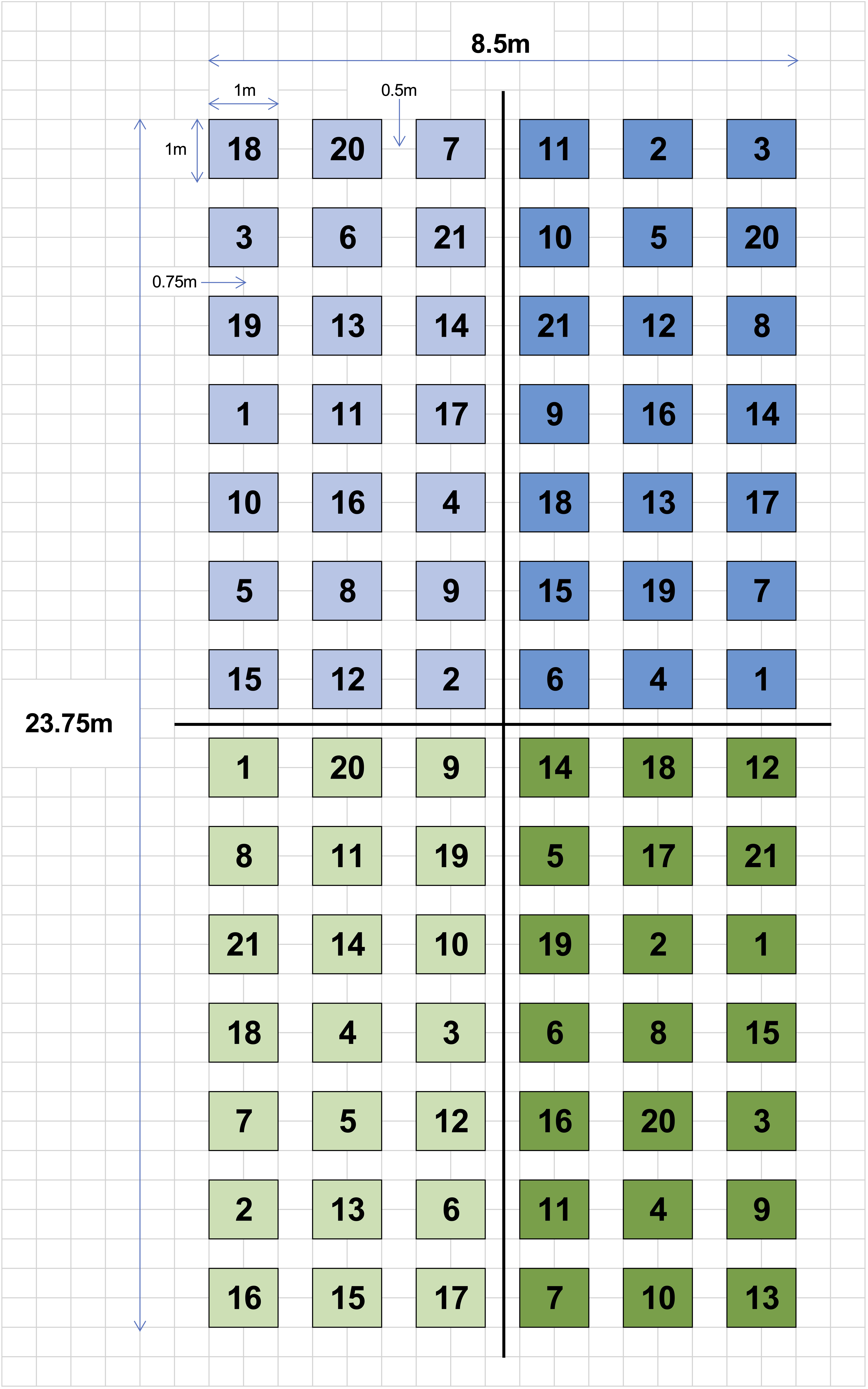}\\ 
\end{figure}

The responses measure the level of aphid colonisation of individual plants, aggregated within each plot. Farm operations suggest the need for additional blocking structures beyond the relatively compact complete replicate blocks of experimental units (21 plots in 7 rows and 3 columns, nested within 2 superrows and 2 supercolumns so that each superrow by supercolumn combination formed a replicate block, see Figure \ref{fig1}). The two farm operations potentially influencing the responses in a systematic way are the drilling of the plots, where a column of 14 plots will be drilled in one pass of the machinery, and the application of any crop protection or nutritional sprays, applied by boom sprayer with rows of 6 plots across the design sprayed simultaneously. These two constraints justify the use of a conventional nested row-column design allowing for variation between rows of 3 plots and columns of 7 plots nested within each complete block, but also suggest a need for blocking structures along long rows of 6 plots and long columns of 14 plots across complete blocks.

Figure \ref{fig1} depicts the physical field layout with the original randomisation of the 21 treatments (lines). The different colours denote the four complete blocks of the nested row-column blocking structure. The design was constructed using the CycDesigN package (https://www.vsni.co.uk/software/cycdesign) as a resolvable row-column $\alpha$-design with rows of 3 plots and columns of 7 plots, latinised by rows (so that any treatment only occurs once in each long row of 6 plots), partially latinised by long columns (but some treatments appear more than once in some long columns of 14 plots) \citep{Williams2013}. Given that for this implementation of the experiment the complete blocks are already defined, we call the design resolved rather than resolvable. The resulting design is therefore a resolved nested row-column design with additional cross-replicate blocking.

The potential interference among plots can be described via a network structure, which performs one of two functions: either capturing the spatial structure to reflect distances between neighbouring plots across space or adjusting for farmer operations. The aim of the motivating experiment was to compare the selected wheat lines with regards to natural cereal aphid colonisation, potentially identifying lines that show some resistance to colonisation. The experimental units correspond to plots (areas of land), which form a network structure pre-specified by the physical arrangement and practical management of the field experiment, so as to address the particular characteristics of the field experiment. The plots constitute the vertices of the network, with adjacencies based on the distances between the plot centroids, allowing for both the spatial separation of plots and the plot size and shape. In particular, the horizontal, vertical and diagonal distances between the centroids are 1.5m, 1.75m and 2.3m respectively (see Figure \ref{fig1}).

 Different causes of interference among neighbouring units (vertices) can result in different specifications of the network structure. Issues to be considered include non-directional or directional interference (e.g.\ impacts of spray drift might be considered as non-directional as wind direction is probably unknown, whereas impacts of shading are probably directional defined by the orientation of the experiment and individual plots), and unweighted or weighted interference effects (e.g.\ weightings might allow for geographical distances or a different impact of neighbouring units at a higher or lower altitude). When dealing with spatial arrangements of experimental units, there are several methods available to specify the neighbour structure. In this case, the experimental units are arranged in a rectangular array, and the physical distances between the units provide a natural basis for calculation of the network structure. We should note, however, that plots are not contiguous but rather they are separated by small distances, with different distances across rows and along columns. 

The network structure can accommodate other factors which could have an impact on the analysis of the field experiment related to farm management, e.g. sets of plots that are sprayed simultaneously or cultivated in a particular order. Alternatively such factors could be incorporated as part of the blocking structure, if there is potential systematic variation due to groups of plots being managed simultaneously. We consider the effects due to the network structure separately to the different levels of blocking complexity and additional to the particular treatment allocation.  

The specification of the network structure may not be straightforward and this structure will likely be a proxy for the actual dynamics (e.g. farmer operations) and interactions that take place between the plots. We assume that the network structure is pre-specified (non-stochastic) and appropriately captures the observed or potential associations among the plots. In common with other features of the design of experiments, such as the blocking structure, subject matter expertise is needed to define an appropriate network structure for a given experiment.

\begin{figure}[h!]
\centering
\caption{Different connectivity graphs: $\mathcal{G}_1$ (\textit{left}) and $\mathcal{G}_2$ (\textit{right})}\label{fig:adjacency}
    \includegraphics[width=.95\textwidth]{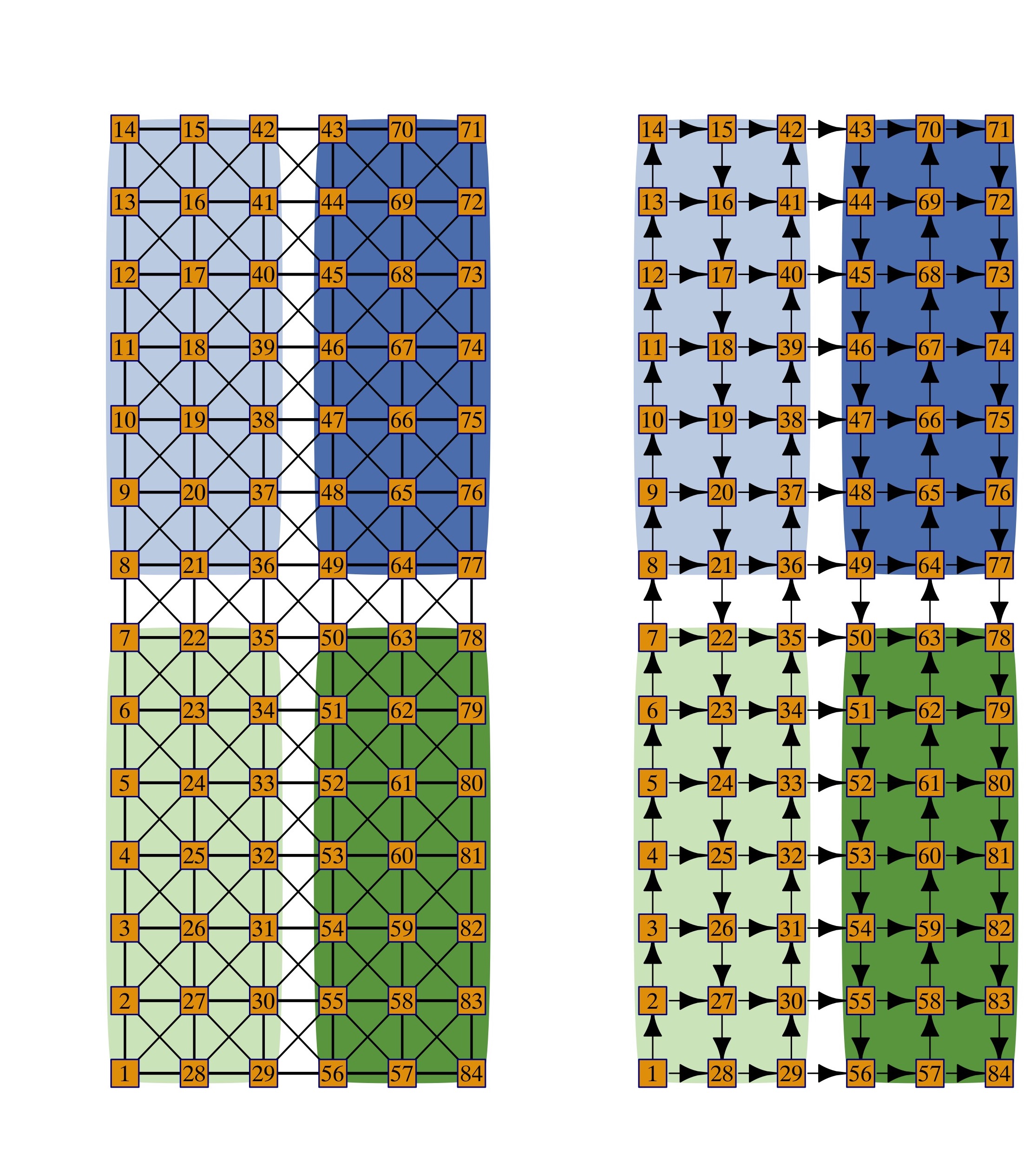}\\
\end{figure}

We consider two choices for this network structure as shown in Figure \ref{fig:adjacency}. These network structure specifications can be used for controlling for unwanted geographical differences in the site and causes of variation incurred from the farm operations and could be appropriately altered if needed for capturing the spatial patterns even of an irregular arrangement of plots. Firstly, we consider the direct competitive effects of the treatments applied to the immediate neighbours that are vertically, diagonally or horizontally connected. The second choice for the adjacency matrix relates to the farmer operations.  

More specifically, for our first specification the adjacency matrix is weighted, with weights based on the inverse spatial distance between plots (centroids) including plots that are very close diagonally to one another, while for the second specification the adjacency matrix is constructed without imposing any weights and it is related to the farm operations and specifically the directions of drilling and spraying applications implemented by the farmer. For brevity we refer to these adjacency specifications as King's case (as in chess) and Farmer's case respectively. 

These adjacency matrices have been specifically chosen to represent two common forms of neighbour effects that might occur across agricultural field experiments, accounting for important components of the trial layout that might otherwise introduce a bias to the results and therefore to the conclusions drawn by the experimenter.  We might expect direct interference effects to stem from the treatments applied to all adjacent plots, and indirect interference effects because of the order of particular farm management operations. This led us to consider a weighted and undirected network for the direct interference effects, with the weights being the reciprocals of the distances between the plot centroids, and a directed but unweighted network for the indirect effects, where the directions indicate the order in which farm management operations are planned to be applied. Section \ref{sec:6} explores the optimality of the designs for these two alternative scenarios of treatment interference among neighbouring plots.

From an analysis of the data collected from the motivating agricultural field experiment there was evidence that the network effects are important despite the highly complex blocking structure (for removing spatial variation). In particular, Tables \ref{tab:BRCM1}--\ref{tab:BRCNM_Farmers1} report the results of nested model comparisons accounting or not for network effects, including two different comparisons for each of the network effect specifications. Fitting the model with network effects and then adding direct effects adjusting for network effects (Comparison 1 in Tables \ref{tab:BRCNM_Kings1} and  \ref{tab:BRCNM_Farmers1}) aligns with the design perspective of estimating direct effects in the presence of nuisance network effects, while fitting the model with direct effects and then adding network effects adjusting for direct effects (Comparison 2) aligns with the modelling perspective of understanding what is the added contribution of network effects. Direct treatment and network effects were modelled as fixed effects and block effects as random. Satterthwaite's approximation method \citep{Satterthwaite1946, FaiandCornelius1996} was used for the denominator degrees of freedom. The estimated residual variances for the models without and with network effects are 0.71, 0.50 (King's case) and 0.59 (Farmer's case) respectively. We can observe that the network effects are significant and that the background variation was partly explained by allowing for these network effects, which is a justification of the need for designing the experiment accounting for the network effects in addition to the complex blocking structure. A more detailed discussion can be found in the Appendix.

\begin{table}[ht!]
\centering
\caption{Analysis without network effects}
\begin{tabular}{lrrrrrr}
  \hline
 & Sum Sq & Mean Sq & NumDF & DenDF & F-value & p-value \\ 
  \hline
Variety & 31.70 & 1.59 & 20.00 & 57.87 & 2.23 & 0.0095 \\ 
   \hline
\end{tabular}
\label{tab:BRCM1}
\end{table}

\begin{table}[ht!]
\centering
\caption{Analysis with network effects for the King's case}
\begin{tabular}{lrrrrrrrr}
\hline
& & Sum Sq & Mean Sq & NumDF & DenDF & F-value & p-value\\ 
\hline
\bf{Comparison 1} & Network effect  & 32.58 & 1.63 & 20.00 & 31.76 & 3.24 & 0.0015\\ 
 & Variety & 19.34 & 0.97 & 20.00 & 35.48 & 1.92 & 0.0437\\ 
\\\hline\\
\bf{Comparison 2} &Variety & 32.20 & 1.61 & 20.00 & 35.85 & 3.20 & 0.0012\\ 
& Network effect & 20.41 & 1.02 & 20.00 & 32.09 & 2.03 & 0.0361 \\ 
\hline
\end{tabular}
\label{tab:BRCNM_Kings1}
\end{table}

\begin{table}[ht!]
\centering
\caption{Analysis with network effects for the Farmer's case}
\begin{tabular}{lrrrrrrr}
 \hline
 & & Sum Sq & Mean Sq & NumDF & DenDF & F-value & p-value \\ 
  \hline
\bf{Comparison 1} & Network effect  & 19.73 & 0.99 & 20.00 & 32.65 & 1.68 & 0.0911\\ 
  & Variety & 23.85 & 1.19 & 20.00 & 33.43 & 2.03 & 0.0340\\ 
\\\hline\\
\bf{Comparison 2} & Variety & 26.80 & 1.34 & 20.00 & 33.24 & 2.28 & 0.0170\\ 
&Network effect & 15.76 & 0.79 & 20.00 & 34.05 & 1.34 & 0.2188\\ 
\end{tabular}
\label{tab:BRCNM_Farmers1}
\end{table}

\section{The design and the model}
\label{sec:4} 

We assume a general row-column structure and then we group rows into superrows (as shown by the green and blue shading in Figure \ref{fig1}) and columns into supercolumns (as shown by the light and dark shadings in Figure \ref{fig1}). We have (superrows/rows)$\times$(supercolumns/columns), where superrows and supercolumns correspond to sets of adjacent rows and sets of adjacent columns respectively (see, e.g., \citealp{Bailey2008, Wingen2014}). Moreover, there is assumed to be an underlying network structure governing the experimental units (plots) which is represented by means of a graph $\mathcal{G}=(\mathcal{V},\mathcal{E})$, with vertex set $\mathcal{V}$ (of size $n$) and edge set $\mathcal{E}$ (of size $l$). The adjacency matrix of a graph is an $n\times n$ matrix $A=\left[A_{jh}\right]$ with $j,h \in \mathcal{V}$, which is a compact way to represent the connectivity structure. The elements of the matrix indicate whether pairs of vertices are adjacent or not in the graph. We aim to improve accuracy as well as the precision of the experiment by controlling for heterogeneity due to multiple sources and by adjusting for the interference between neighbouring units. Note that we are designing the experiment on the network with respect to fixed effects for each of the blocking model terms. If block labels are properly randomised to blocks, it is reasonable to analyse these trials with recovery of inter-block information by using random effects for blocks. However, we will develop a design criterion using the model with fixed block effects. Although the model with random block effects may well be better for analysis, the variances of estimated treatment parameters depend on the ratio of between- and within-block variances. Since this ratio is unknown, it is safest to design for the worst case, i.e., that which leads to the largest variances of estimated treatment parameters. This occurs when the between-block variance tends to infinity, which is equivalent to the fixed blocks case.

\begin{figure}[h!]
\centering
\caption{Hasse diagram of the unit structure of the design. Each node has two numbers: the number of levels of the corresponding blocking factor and the corresponding degrees of freedom (in brackets), obtained by subtracting the degrees of freedom for higher factors from the number of levels of the factor under consideration. R, C, r and c represent supperrows, suppercolumns, rows and columns respectively. }\label{fig:hasse_RCD}
    \includegraphics[width=0.45\textwidth]{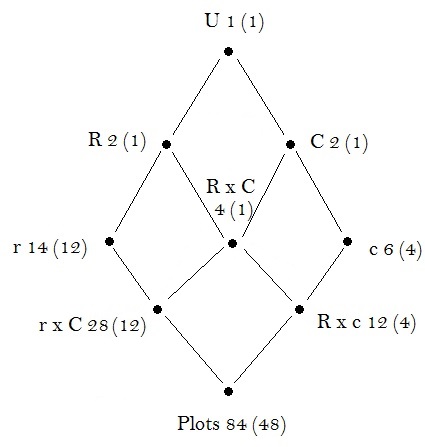}\\
\end{figure}

Returning to the motivating agricultural experiment, the entire experiment is broken down into superrows and supercolumns of lengths $b_1=b_2=2$ (see Figure \ref{fig1}, the first superrow contains the blue-shaded plots and the second superrow contains the green shaded plots, and the first supercolumn contains the light shaded plots and the second supercolumn contains the dark shaded plots). The superrow by supercolumn combinations are equal to the $\kappa=4$ blocks, and the complete array can also be broken down to $\kappa_1=14$ rows each containing 6 plots and $\kappa_2=6$ columns each containing 14 plots. The Hasse diagram (see, e.g., \citealp[Ch.10.4]{Bailey2008}) in Figure~\ref{fig:hasse_RCD} describes the unit structure for this experiment with the corresponding degrees of freedom in each stratum. Recall that for the original design of the trial, the management operations of drilling and spray applications are done column-by-column and row-by-row respectively, so that it is assumed that the effects of these processes will be confounded with the positional effects (i.e. the row, column, superrow and supercolumn effects and interactions among these terms).

Let $y_{ikgh}$ denote the response from the experimental unit in the $g$-th row and $h$-th column within the $i$-th superrow and $k$-th supercolumn. The quadruple $(i,k,g,h)$, where $i=1,2,\ldots,b_1$; $k= 1,2,\ldots,b_2$; $g=1,2,\ldots,\kappa_1$; and $h=1,2,\ldots,\kappa_2$, identifies the experimental unit which corresponds to the vertex $v \in \left\{1,\ldots,n\right\}$ and $t(ikgh)$  $\in \left\{1,\ldots,m\right\}$ corresponds to the treatment applied to unit $(i,k,g,h)$. We should also note that the identity of every vertex in the network has been fixed by assigning a unique label to each one (given that a labelling is arbitrary and every choice leads to an equivalent description of the same network space). The most complex model we might consider, incorporating blocks (combinations of superrows and supercolumns), rows and columns within superrows and supercolumns respectively, and network effects. Hence the Block Row-Column Network Model (BRCNM), which is an extension of the block network model \citep{Koutraetal2021}, is 

\begin{align}
\mbox{BRCNM:} \; y_{ikgh} &=\mu+\tau_{t(ikgh)}+\sum_{i'=1}^{b_1}\sum_{k'=1}^{b_2}\sum_{g'=1}^{\kappa_1}\sum_{h'=1}^{\kappa_2} A_{\left\{ikgh,i'k'g'h'\right\}} \gamma_{t(i'k'g'h')} \nonumber \\
&+R_i+C_k+(RC)_{ik}+r_{ig}+c_{kh}+{\left(rC\right)}_{ikg}+{\left(Rc\right)}_{ikh} +\epsilon_{ikgh}  , \label{BRCNM} 
\end{align}	

\vspace{.5cm}

\noindent where $\mu$ denotes the overall mean effect, $\tau_{t(ijkl)}$ is the (direct) treatment effect, $R_i$ and $C_j$ denote the $i$-th and $j$-th superrow and supercolumn effects respectively, while $r_{ig}$ and $c_{kh}$ denote the row and column effects nested within the superrows and supercolumns respectively, $(RC)_{ik}$ denotes the interaction effects of superrows and supercolumns, ${\left(rC\right)}_{ikg}$ denotes the interaction effects of rows and supercolumns, and similarly ${\left(Rc\right)}_{ikh}$ denotes the interaction effects of columns and superrows. The adjacency matrix $A_{\left\{ikgh,i'k'g'h'\right\}}$ indicates connections between the units $(i,k,g,h)$ and $(i',k',g',h')$ of the weighted and/or directed graph and $\gamma_{t(i'k'g'h')}$ is the network effect of the treatment $t(i'k'g'h')$ applied to the connected unit $(i',k',g',h')$ when there is connection between the two experimental units defined by these indices (neighbour or indirect treatment effect or interference effect). By convention the diagonal elements of the adjacency matrix are all set to zero for avoiding self loops. The $\epsilon_{ikgh}$ are assumed to be independent random variables, each with $E(\epsilon_{ikgh}) =0$ and $E({\epsilon_{ikgh}}^2 )=\sigma^2$.

The model in matrix notation can be written as
\begin{align}
\e{\boldsymbol{y}}&=\mu \boldsymbol{1}+X_{\tau} \boldsymbol{\tau}+A X_\gamma\boldsymbol{\gamma}+X_R \boldsymbol{R}+X_C\boldsymbol{C}+X_{RC}(\boldsymbol{R}\boldsymbol{C})+X_r \boldsymbol{r}+X_c \boldsymbol{c}+X_{r C}(\boldsymbol{r}\boldsymbol{C})+X_{Rc}(\boldsymbol{R}\boldsymbol{c}), \nonumber 
\end{align}

\noindent where $\boldsymbol{\tau}$, $\boldsymbol{\gamma}$, $\boldsymbol{R}$, $\boldsymbol{C}$, $(\boldsymbol{R}\boldsymbol{C})$, $\boldsymbol{r}$, $\boldsymbol{c}$, $(\boldsymbol{r}\boldsymbol{C})$ and $(\boldsymbol{R}\boldsymbol{c})$ denote the vectors of treatment, network, superrow, supercolumn, superrow$\times$supercolumn interaction, row, column, row$\times$supercolumn interaction and superrow$\times$column interaction effects respectively. The model matrices $X_\tau$,  $X_\gamma$, $X_R$, $X_C$, $X_{RC}$, $X_r$,  $X_c$, $X_{r C}$ and $X_{Rc}$ represent the treatments, network effects, superrows, supercolumns, superrow$\times$ supercolumn interactions, rows, columns, row$\times$supercolumn interaction and superrow$\times$column interaction respectively. 

Thus the information matrix for this model is $M=X^TX$, where 
\[X=\left(\boldsymbol{1} \; X_\tau \; A X_\gamma  \; X_R \; X_C \; X_{RC} \; X_r \; X_c \; X_{r C} \;X_{Rc}  \right).\] 

The information matrix has the form
\[M=
{\left( \begin{array}{ccccc}
n & \textbf{1}^T {X_\tau} &  \textbf{1}^T A {X_\gamma} &\ldots &\textbf{1}^T X_{Rc} \\
{X_\tau}^T\textbf{1} & {X_\tau}^T {X_\tau} & {X_\tau}^T A {X_\gamma}& \ldots &{X_\tau}^T X_{Rc} \\
{X_\tau}^T A \textbf{1} & {X_\tau}^T A {X_\tau} & {X_\tau}^T A^2 {X_\gamma}& \ldots &{X_\tau}^T A X_{Rc} \\

\vdots & \cdots &\vdots& \ddots & \vdots \\
X_{Rc}^T \textbf{1}  & X_{Rc}^T {X_\tau} & X_{Rc}^T A {X_\gamma}&\ldots &X_{Rc}^T X_{Rc} 
\end{array} \right).
}
\] 

We are interested in obtaining the least squares estimators of the pairwise differences of the treatment effects, $\widehat{\tau_s-\tau_{s'}}$ with $s,s' \in \left\{1,\ldots,m\right\}$. The design performance will be assessed via the $A_s$-optimality criterion, which minimises the average variance of all pairwise differences of treatment comparisons 
\[
\frac{2}{m(m-1)} \sum_{s=1}^{m-1}\sum_{s'=s+1}^m var(\widehat{\tau_s-\tau_{s'}}). \nonumber
\]
This is proportional to 
\[
\phi =\sum_{v=2}^{m} \sum_{h=v+1}^{m+1} \boldsymbol{s}^T(v,h) M^{-} \boldsymbol{s}(v,h)\,,
\]

\noindent where $\boldsymbol{s}(v,h)$ is a contrast vector formed of zeroes of appropriate dimension, except the $v$ and $h$ elements which are $1$ and $-1$ respectively, corresponding to the particular treatments we wish to compare in a given contrast. The $M=M(\xi)$ corresponds to the information matrix and $\xi$ is a design chosen from $\Xi$ the set of all possible designs, where the design is a choice of treatment assignments to the experimental units that correspond to the vertices of the network. The $M^{-}$ is a generalised inverse of the information matrix $M$ \citep[Ch.9]{Harville1997}, which is pre- and post-multiplied by this contrast vector in the summation that defines the optimality criterion $\phi$ across the 210 pairwise comparisons between varieties (treatments) in this experiment. Our interest is in the variance of estimable contrasts, which is invariant with respect to the choice of the generalised inverse. Thus the minimisation of the criterion function leads to an $A_s$-optimal design and the minimum value of $\phi$ is the optimal function value. \cite{Koutraetal2021} provides a more thorough description of the formulation of the optimality criterion but for a simpler blocking structure. 

For each potential model, the optimal design will be obtained following an optimisation algorithm that is described in the Appendix, by generating an initial treatment arrangement with some particular properties, e.g.\ resolved row-column design, and making pairwise interchanges of treatments between plots (restricting these interchanges to maintain the overall properties where appropriate, e.g. only making interchanges within blocks for a resolved design). We have chosen a fairly standard exchange type algorithm because its flexibility can be used for any design structure and it seems to work adequately for the problems of interest. However other methods can be used such as tabu search \citep{Glover1989}, or simulating annealing \citep{Kirkpatrick1983}.

For our comparisons, we consider models that are special cases of the BRCNM. In particular:
\begin{align*}
\mbox{CRM:} \; \; \e{\boldsymbol{y}}&=\mu \boldsymbol{1}+X_{\tau} \boldsymbol{\tau}\,, \\[1.5ex]
\mbox{RBM:} \; \; \e{\boldsymbol{y}}&=\mu \boldsymbol{1}+X_{\tau} \boldsymbol{\tau}+X_R \boldsymbol{R}+X_C\boldsymbol{C}+X_{RC}(\boldsymbol{R}\boldsymbol{C})\,, \\[1.5ex]
\mbox{RCM:} \; \;  \e{\boldsymbol{y}}&=\mu \boldsymbol{1}+X_{\tau} \boldsymbol{\tau}+X_r \boldsymbol{r}+X_c \boldsymbol{c}\,, \\[1.5ex]
\mbox{BRCM:} \; \; \e{\boldsymbol{y}}&=\mu \boldsymbol{1}+X_{\tau} \boldsymbol{\tau}+X_R \boldsymbol{R}+X_C\boldsymbol{C}+X_{RC}(\boldsymbol{R}\boldsymbol{C}) +X_r \boldsymbol{r}+X_c \boldsymbol{c}+X_{r C}(\boldsymbol{r}\boldsymbol{C})+X_{Rc}(\boldsymbol{R}\boldsymbol{c})\,, \\[1.5ex]
\mbox{LNM:} \; \; \e{\boldsymbol{y}}&=\mu \boldsymbol{1}+X_{\tau} \boldsymbol{\tau}+A X_\gamma\boldsymbol{\gamma}\,, \\[1.5ex] 
\mbox{BNM:} \; \; \e{\boldsymbol{y}}&=\mu \boldsymbol{1}+X_{\tau} \boldsymbol{\tau}+A X_\gamma\boldsymbol{\gamma}+X_R \boldsymbol{R}+X_C\boldsymbol{C}+X_{RC}(\boldsymbol{R}\boldsymbol{C})\,, \\[1.5ex]
\mbox{RCNM:}\;  \; \e{\boldsymbol{y}}&=\mu \boldsymbol{1}+X_{\tau} \boldsymbol{\tau}+A X_\gamma\boldsymbol{\gamma}+X_r \boldsymbol{r}+X_c \boldsymbol{c}\,. 
\end{align*}

We consider the standard treatment models derived from the simplest randomisation schemes, the Completely Randomised Model (CRM) and the Randomised (complete) Block Model (RBM). The corresponding designs for these models are the simplest forms of designs to compare different treatments by randomly assigning them to experimental units (the Completely Randomised Design (CRD) and the Randomised (complete) Block Design where treatments are additionally arranged in, potentially resolvable, blocks). If instead of a simple blocking structure, we allow for experimental units being in a two-dimensional arrangement of rows and columns then we have the Row-Column Model (RCM) and if, additionally, we consider the row and column effects to be nested within blocks we have the Block Row-Column Model (BRCM). Extending these four models (CRM, RBM, RCM, BRCM) to include a network model term to capture the connections among units lead to the Linear Network effects Model (LNM) (introduced in \citealp{Parkeretal2017}), the Block Network Model (BNM) (introduced in \citealp{Koutraetal2021}), the Row-Column Network Model (RCNM) and the Block Row-Column Network Model (BRCNM). The models are functions of the network effects, the treatment (and blocking) factors, plus the error terms. We assume that, in all cases, the errors are independent and random with zero mean and constant variance. Our interest lies in comparing designs under the same model, making $\sigma^2$ redundant as it is the same for all proposed designs under the same model.    

\section{Comparison of designs} 
\label{sec:6}

In this section comparisons are provided of optimal designs for estimating the treatment differences under different models for the two different pre-specified adjacency matrices. We consider designs with unequal and equal replication restricting the latter additionally for resolvability. This allows us to measure the efficiency loss by imposing additional restrictions on the randomisation process: for instance comparing a resolved row-column design with network effects and equal replication to a row-column design with network effects, which is not restricted to be resolved or equally replicated. In doing so we obtain the optimality function values of each of the optimal designs with respect to the different models. 

Designs for the corresponding models we compare are labelled as CRD (for the Completely Randomised Design), RBD (for the Randomised Block Design), RCD (for the Row-Column Design), BRCD (for the nested Block Row-Column Design), LND (for the network design under the Linear Network effects Model), BND (for the Block Network Design), RCND (for the Row-Column Network Design), and BRCND (for the Block Row-Column Network Design). We compare the performance of two designs using their \textit{relative $A_s$-efficiency}, which, with respect to the objective function $\phi$ of a design $\xi_2$ compared with a design $\xi_1$, is given by $\mbox{Eff}(\xi_1,\xi_2)=\phi(\xi_1)/\phi(\xi_2)$. We can also define the $A_s$-\textit{efficiency} of a design $\xi$ as $\mbox{Eff}(\xi)=\mbox{Eff}(\xi^*,\xi)$, where $\xi^*$ is an $A_s$-optimal design. The relative efficiencies for different designs are in the context of assuming a particular model, with most interest in comparing alternative designs with the optimal design for each model. We return to these definitions when assessing the performance of the optimal designs for different models for the two different network specifications.

We investigate the benefits of imposing equal replication and/or resolvability, wherever these restrictions can be imposed under a class of designs. In general, we get a loss of efficiency both with adding restrictions to the randomisation process and also by adding restrictions to the relative replications -- so when we assume a more complex model, an optimal design for a simpler model will be less $A_s$-efficient.

As will be seen later, the arguments against imposing equal replication and resolvability are weaker if we have already included network effects in the design, in the sense that we are not losing much in efficiency by imposing further restrictions on the optimisation process. This is an important observation given that algorithmically it is better to impose more restrictions since we reduce the design space leading to a faster convergence to an efficient design. We should also remember that the algorithm presented in the Appendix has been adjusted appropriately for producing each class of designs by imposing additional restrictions. For the unrestricted case the algorithm is an interchange-exchange algorithm running two nested computations sequentially: visiting each unit in order and freely allowing for an exchange of the treatment with any of the listed competitive treatments and then interchanging those treatments until we reach convergence. Note that the optimal designs for the first step under CRM, RBM, RCM and BRCM always have equal replication, without imposing any constraint. Note also that we always assume that competing block designs have the same block partition. 

\bigskip
\textbf{King's case ($\mathcal{G}_1$)}
\bigskip

We first consider the King's case where the adjacency matrix represents the direct effects of treatments in up to eight neighbouring plots (fewer for the edge plots). Thus it has weights corresponding to the reciprocal of the distances between the plot centroids. The optimality function values for each obtained design (labelling the rows of the table) under the different models (labelling the columns of the table) are given in Table \ref{tab:phi_kings}. The criterion values of the optimal designs for each model are highlighted in bold. Recall that the smaller the criterion value the better the design is. Note that the optimal designs under the standard models have been chosen arbitrarily (multiple designs will be optimal where no network effects are included). 

\begin{table}[]

\caption{King's Case: values of the objective function $\phi$ for the optimal designs under the different models (smaller $\phi$ the better the design is)}
\centering
\begin{tabular}{lllllllll}
                    &     &     &     &      &     &     &      &       \\
  & \multicolumn{8}{c}{\textbf{Models}}     \\
  & \multicolumn{8}{c}{}     \\
\textbf{Classes of  designs}  & \multicolumn{8}{c}{}     \\

                    & CRM & RBM & RCM & BRCM & LNM & BNM & RCNM & BRCNM \\
\textbf{Unrestricted}  		  &     &     &     &      &     &     &      &       \\
CRD                 & \textbf{105} & 111   & 146   & 159   & 192  & 238   & 682     & 642  \\
RBD                 & 105 & \textbf{105} & 129   & 142    & 228   & 249   & 547     & 589   \\
RCD                 & 105 & 108  & \textbf{126} & 143  & 199 & 239 &  552  & 550  \\
BRCD1               & 105 & 106 & 140 & \textbf{126}  & 190 & 210 & 502  & 499   \\
                             &     &     &     &      &     &     &      &       \\
LND1                & 106 & 110 & 149 & 153  & \textbf{130} & 158 & 550  & 549   \\
BND1                & 109 & 112 & 150 & 159  & 137 & \textbf{144} & 622  & 506   \\
RCND1               & 108 & 112 & 141 & 153  & 152 & 167 & \textbf{239}  & 353   \\
BRCND1              & 108 & 114 & 148 & 147  & 155 & 174 & 373  & \textbf{254}   \\
                    &     &     &     &      &     &     &      &       \\\hdashline\\
\textbf{Equal-replicated}   &     &     &     &      &     &     &      &       \\
LND2                & 105 & 106 & 145 & 145  & \textbf{140} & 155 & 622  & 625   \\
BND2                & 105 & 105 & 145 & 142  & 141 & \textbf{146} & 604  & 570   \\
RCND2               & 105 & 106 & 134 & 142  & 178 & 191& \textbf{311}  & 512   \\
BRCND2              & 105 & 105 & 144 & 144  & 163 & 176 & 401  & \textbf{315}   \\

                    &     &     &     &      &     &     &      &       \\\hdashline \\
\textbf{Resolved}       &     &     &     &      &     &     &      &       \\
BND3                & 105 & 105 & 143 & 142  & 144 & \textbf{148} & 625  & 533  \\
BRCND3              & 105 & 105 & 140 & 146  & 167 & 174 & 474  & \textbf{318}  \\ \\ \\
&     &     &     &      &     &     &      &       \\\hline\\
\textbf{Implemented design}                &     &     &     &      &     &     &      &       \\
$\alpha$-RCD  & 105 & 105 & 132 & 137  & 179 & 191 & 554  & 513   \end{tabular}
\label{tab:phi_kings}
\end{table} 
       
Focusing on the last column where the true model is assumed to be the BRCNM, we can see that all standard randomised designs without network effects (which we will refer to as non-network designs) perform poorly with approximate $A_s$-efficiencies of $40\% \; (=254/642)$, $43\% \; (=254/589)$, $46\% \; (=254/550)$ and $51\% \; (=254/499)$ for the optimal CRD, RBD, RCD and BRCD respectively. Moreover, when we account for block effects, in addition to network effects, the designs perform slightly better than when ignoring them; efficiency increases to $48\%$ ($= 254/506$ for BND1) compared with $46\%$ ($= 254/549$ for LND1). Accounting additionally for the row and column effects (in RCND1) the design is $71\% \; (= 254/353$) efficient. Thus we can see that with respect to this criterion all designs perform poorly under BRCNM, which means that if we strongly believe that all these effects are present, we should account for them in the design. If we do not believe that there are block effects, by including them the efficiency decreases to $88\% \; (=239/373)$ (see $\phi_{\tiny\mbox{BRCND1}}$ under RCNM). 

As we can observe from the class of designs with equal replication, if we impose this constraint we lose efficiency compared to not doing so. It is interesting to note that the differences in the efficiencies for designs that account for the network structure compared to those without are relatively small. It can also be seen that there is modest loss of efficiency for using a more complex design approach. For example, LND1, BND1, RCND1 and BRCND1 are $85\% \; (=126/149)$, $84\%\; (=126/150)$, $90\% \; (=126/141)$ and $85\% \; (=126/148)$ efficient, under the Row-Column Model (RCM). Note also that the BND1 performs almost as well as the LND1 (the same holds for BND2 compared to LND2) under LNM indicating that we do not do much worse by including block effects in the network model in terms of the design efficiencies.  Additionally forcing resolvability we obtain the third class of optimal designs that perform similarly to those found under the constraint of equal replication. 

Another observation from the results is that some of the biggest differences occur when there are network effects with and without the row-column structure. In particular, when the unit structure is left out from the design, the efficiency drops. Also including the row-column structure nested within blocks in the design, there is a smaller but still significant reduction in efficiency possibly due to the additional imposed structure conflicting with the network structure.

\bigskip
\textbf{Farmer's case ($\mathcal{G}_2$)}
\bigskip

At this point, we focus on the adjacency matrix related to the farmer operations. The optimality function values for the optimal designs under each of the models are shown in Table \ref{tab:phi_farmer}. We can obtain the $A_s$-efficiencies of each design with respect to the optimal design. We note that in the first class of candidate designs where we allow for non-resolvability and unequal replication, the optimal designs are all equally replicated (including those accounting for network effects). 

We see that the results here follow similar patterns to the results in the King's case. One difference stemming from the different network specification is that the optimal function values of the non-network designs are slightly better than before. Assuming, for instance, that BRCNM is true, the non-network designs have approximate $A_s$-efficiencies of $51\% \; (=174/343)$, $64\% \; (=174/273)$, $57\% \; (=174/306)$ and $68\% \; (=174/255)$ for the CRD, RBD, RCD and BRCD respectively. Accounting additionally for the network effects, the optimal designs are $65\%  \; (=174/266)$, $78\%  \; (=174/223)$ and $78\%  \; (=174/222)$ efficient for the LND4, BND4 and RCND4 respectively, implying that accounting for the block effects is as good as accounting for the row and column effects. When we additionally restrict for resolvability we lose less than $10\%$  in efficiency (e.g.\ $92\%=174/189$ for BRCND5 and $97\%=132/136$ for BND5 efficient relative to the non-resolved designs), but the BND5 compared to the BRCND5 performs very similarly in terms of efficiency under all models. 

In general, we can infer that when we believe that there may be important spillover (neighbour) effects due to a structure governing the plots under experimentation, it is sensible to incorporate them in the model. For this second network specification (the Farmer's case) accounting for the network effects even when we question their true existence does not do much harm in the design efficiencies, which means that we are better off accounting for network effects than ignoring them.

\begin{table}[h!]

\caption{ Farmer's Case: values of the objective function $\phi$ for the optimal designs under the different models (smaller $\phi$ the better the design is)}
\centering
\begin{tabular}{lllllllll}
                    &     &     &     &      &     &     &      &       \\
  & \multicolumn{8}{c}{\textbf{Models}}     \\
 & \multicolumn{8}{c}{}     \\
\textbf{Classes of  designs} & \multicolumn{8}{c}{}     \\

                    & CRM & RBM & RCM & BRCM & LNM & BNM & RCNM & BRCNM \\
\textbf{Unrestricted}  		  &     &     &     &      &     &     &      &       \\
CRD                          & \textbf{105} & 111 & 146 & 159  & 164 & 183 & 306  & 343   \\
RBD                          & 105 & \textbf{105} & 129 & 142  & 159 & 161 & 249  & 273   \\
RCD                          & 105 & 108 & \textbf{126} & 143  & 167 & 176 & 249  & 306   \\
BRCD                         & 105 & 106 & 140 & \textbf{126}  & 162 & 168 & 275  & 255   \\
                             &     &     &     &      &     &     &      &       \\
LND4                         & 105 & 110 & 140 & 143  & \textbf{130} & 143 & 223  & 266   \\
BND4                         & 105 & 106 & 141 & 145  & 130 & \textbf{132} & 240  & 223   \\
RCND4                       & 105 & 108 & 131 & 142  & 134 & 141 & \textbf{170}  & 222   \\
BRCND4                       & 105 & 107 & 138 & 131  & 134 & 138 & 233  & \textbf{174}   \\
                             &     &     &     &      &     &     &      &       \\\hdashline\\
\textbf{Resolved}                &     &     &     &      &     &     &      &       \\
BND5                         & 105 & 106 & 146 & 148  & 135 & \textbf{136} & 237  & 242   \\
BRCND5                       & 105 & 106 & 144 & 133  & 139 & 140 & 244  & \textbf{189}  \\
&     &     &     &      &     &     &      &       \\\hline\\
\textbf{Implemented design}                &     &     &     &      &     &     &      &       \\
$\alpha$-RCD  & 105 & 105 & 132 & 137  & 153 & 155 & 258  & 282   
\end{tabular}
\label{tab:phi_farmer}
\end{table}

\newpage
\textbf{Implemented design for motivating example}
\bigskip

We obtain the objective function values for the resolved $\alpha$-RCD, the design actually used for our motivating example, under the different models. The design is shown in Figure~\ref{fig1}. By ignoring the network effects under the assumption that the network structure exists, we observe that the design efficiency, with respect to the resolved BRCND3,  is around $62\% \; (=318/513) $ for the King's case (see Table \ref{tab:phi_kings}). Likewise with respect to the resolved BRCND5, the design efficiency is around $67\% \; (=189/282)$ for the Farmer's case (see Table \ref{tab:phi_farmer}).  This implies that we considerably increase the design efficiency if network effects are both necessary and accounted for in the modelling.

In this section we showed that optimal designs that account for network effects outperformed conventional non-network designs in terms of efficiency when network effects are present. There is also no significant loss of efficiency in assuming a more complex model, when designing the experiment as the optimal network design is about $2\%$ to  $15\%$ less efficient as the standard non-networked design when the network effects are ignored. The loss becomes higher with more complex blocking structure as the randomisation of the allocation of treatments to units matter more.

In the resulting optimal network designs, it was noticed that each replicate of every treatment was close to at least one replicate of all the other treatments, a desirable feature previously highlighted by \cite{Freeman1979} for the case of row-column designs. Also, pairs of closely connected units tend to receive the same treatment, also observed by \citet{Parkeretal2017} and \citet{Koutraetal2021}. The optimal designs can be found in the Supplementary Material.

\section{Discussion}\label{sec:8}

In this study, we attempted to control for the potential variation and bias resulting from treatment interference or farm operations in agricultural field experiments through incorporating network effects in order to improve the precision of treatment comparisons. We show that optimal designs with network effects outperform conventional designs in terms of efficiency, where there is strong expectation of neighbour or network effects. This approach is especially effective when there is good information about potential effects for example associated with the size of effects or the distance of neighbours. Including network effects that might be important is better than ignoring them and still the resulting optimal designs perform well under the conventional models CRM, RBM, RCM etc. Also, by not taking into account network effects in our design, we produce an experiment which can have higher variance than necessary but also biased treatment effect estimators. 

In practice, the adjacency matrix is tailor-made reflecting the suspected underlying interference structure among plots. The choice of this matrix can also address irregular layouts demanding further potential constraints. According to the specific problem at hand, the experimenter should appropriately choose the adjacency matrix, suggest a suitable model to fit and optimise the design for that model for estimating the important parameters of interest. Various alternative specifications of the adjacency matrix might include detailed measurements between neighbouring plots to better reflect the spatial structure, the identification of different field management practices, and the identification of both direct and indirect impacts of treatments on neighbouring plots. In practice, the specification of this network might be a result of the scientific knowledge of the experimenter or elicitation of information from farm managers on the suspected sources of spillover effects based on experience which is likely to strongly affect the differences in treatment effects. A conclusion drawn from the comparison of the optimal designs is that unsystematic designs that ignore network effects are inefficient when network effects are present and may lead to poor results due to additional variability associated with treatment effects. 

Increasing constraints on the funding for agricultural research, particularly field experiments, and an interest in being able to detect smaller and smaller treatment differences as being statistically significant, means that there is increasing demand for the development of statistical design approaches that can lead to the implementation of more efficient field experiments. Although the motivating example had a rectangular array of plots, the adjacency matrix can be easily specified to accommodate any potential inter-plot interference to ultimately reduce spatial and/or other sources of variation. Designs explicitly incorporating neighbour effects overcome these challenges, and the approaches described in this paper, building on the models previously described by \cite{Parkeretal2017} and \cite{Koutraetal2021}, extend the pioneering research of \cite{BesagandKempton1986}. Our approach allows consideration of different types of neighbour effects -- including the direct effects of treatments on neighbouring plots, the indirect effects of the responses to treatments on neighbouring plots, and “nuisance” effects associated with farm management practices -- within a complex nested and/or crossed blocking structure. Thus considering models that combine network and block effects to capture the wide range of “nuisance” and “interference” sources of variability provides an approach to the design of more efficient agricultural field experiments to better address the current challenges of achieving more efficient and sustainable food production with limited resources, through identifying treatments resulting in small incremental improvements. 

The advantage of the suggested design approach is that we can control for the neighbouring environment according to the shapes, sizes, causes, directions and weights of the neighbouring interference. For instance, neighbour effects may depend on the speed and direction of wind or periods in shade, which can result in an appropriate definition of the connectivity matrix imposing suitable weights for the left or right neighbours, etc. This highlights the importance of defining the adjacency matrix based on requirements of the experiment at hand. If the network structure is adequately modelled, this design procedure may be expected to cause an increase in precision of the treatment contrasts. 

This study is intended to encourage the future application for agricultural field experiments of designs assuming models including both network effects and complex blocking structures, to account for the anticipated dependence among neighbouring experimental units alongside those sources of variation conventionally accounted for using blocking.

 \if1\blind
{
\section*{Acknowledgements}
This work was funded by the Economic and Social Research Council (ESRC) and was conducted under an internship scheme at Rothamsted Research in the Applied Statistics Group.
}\fi

\vspace*{-8pt}

\section*{Supplementary Materials}
The designs with network effects discussed in Section ~\ref{sec:6} of the paper are shown in Figures 1 -- 16.

\section*{Appendix}

\subsection*{Optimisation algorithm}

Early algorithms for optimal row-column designs were discussed in detail by \citet{JonesandEccleston1980a,JonesandEccleston1980b}. Their optimal designs were evaluated using the $A_s$-optimality criterion for minimising the sum of the weighted variances of a set of treatment contrasts of interest. Computer algorithms have also been suggested for obtaining row-column designs, for instance, the design generation package ALPHA+ for obtaining $\alpha$-designs developed by \cite{WilliamsandTablbot1993}, described in detail by \cite{NguyenandWilliams1993} and the nested simulated annealing algorithm developed by \cite{JohnandWhitaker1993}. They all use some form of interchange procedure where pairs of treatments are swapped in the design, subject to an iterative improvement procedure with respect to a chosen optimality criterion. To address the problem of the interchange  procedure getting stuck at a local optimum, \cite{NguyenandWilliams1993} suggested repeated runs of the algorithm using different starting designs, and then choosing the best design over all runs. \cite{JohnandWhitaker1993} also addressed this problem by accepting with low probability some randomly chosen interchanges that do not result in an improvement in the chosen optimality criterion.

For the objectives of designing the motivating agricultural field experiment, we describe the algorithm for the most complex case of resolvable blocks, nested row and column effects, and equal replication. This is an interchange algorithm for the construction of efficient resolved row-column designs with network effects. The algorithm begins with the generation of a non-singular design with a crossed blocking structure and a fixed number and sizes of blocks, ensuring that the starting design contains all treatments. The $A_s$-optimality criterion determines the decision rule of either allowing the interchange to occur or leaving the design unchanged. Given that we restrict the design to be resolved, the candidate treatment swaps are restricted to take place only within blocks, accepting those interchanges that improve the criterion value for the overall design. The interchanges of pairs of treatments are chosen systematically within the same block. Thus the algorithm focuses on each block in turn, keeping the treatment allocation fixed in the remaining blocks. The algorithm continues to cycle through the blocks, making interchanges until no beneficial changes can be made in any of the blocks. The steps in the algorithm are as follows:

\begin{itemize}
\item[--] Step 1: Generate a random non-singular resolved row-column design and calculate the optimality criterion function for this starting design.
\item[--] Step 2: Make a pairwise interchange of treatments within the current block keeping the arrangement of the treatment combinations fixed for the remaining blocks. Calculate the optimality criterion function  for the current design that corresponds to that specific interchange. If an interchange improves the criterion value of the overall design, accept it and continue; otherwise, undo the interchange and continue. 
\item[--] Step 3: Repeat Step 2 for for each unit in the block until no further interchanges in the current block result in an improvement (or if the above holds for at least a large number of iterations) then move on to the next block. 
\item[--]  Step 4: Repeat Steps 2 and 3 until a pass through all blocks yields no changes or for a specific number of times.
\item[--]  Step 5: Repeat Steps 1-- 4 for several randomly generated initial designs to overcome the problem of becoming stuck in a local optimum.
\end{itemize}

In Section \ref{sec:6}, we provide different designs adjusting this algorithm appropriately. In particular, we consider two modified algorithms where we drop the resolvability property and/or the constraint of equal replication. In the first case we relax the constraint of having a resolved design with the treatment interchanges occurring between pairs of plots within the same replicate blocks, and let them occur across the whole design (non-resolved but equal replicated), while the second modified algorithm allows for treatment exchanges rather than interchanges to generate non-equireplicate designs. For the exchanges the algorithm moves systematically along all the units and exchanges the treatment with an alternative treatment retaining the exchange if this results in an improvement of the criterion.

\subsection*{Analysis of the motivating agricultural experiment at Rothamsted Research}

We provide the analysis of the experiment run in 2016 to explore the presence of the possible network effects in this type of study. In particular we fit linear mixed effects models using \texttt{lme4} package \citep{lme4} in \texttt{R} \citep{RCore2021} with random block effects to assess differences in total aphid numbers between treatments (wheat lines). We include both direct treatment and network effects as fixed effects and block structures as random effects. We fit both the BRCM and BRCNM (Equation \ref{BRCNM}), where all blocking components are random effects that are induced by restrictions in the randomisation in the unit structure, that is rows, columns, superrows, supercolumns, and their interactions. We fit the models using residual maximum likelihood (REML) and the generalised least squares (GLS) estimation method. Zero variance component estimates were mostly obtained, and that is because there are only very small numbers of degrees of freedom left for estimating them \citep{GilmourandGoos2009}. The estimated residual variances for BRCM and BRCNM (King's and Farmer's cases) are 0.71, 0.50 and 0.59 respectively. Non-orthogonality of the unit structure requires an adjustment to be made to the degrees of freedom, for example using Satterthwaite's approximation method \citep{Satterthwaite1946, FaiandCornelius1996} as implemented in the \texttt{lmerTest} package \citep{lmerTest}.

Visual inspection of residual plots reveal small deviations from homoscedasticity and normality, which were addressed using the Box-Cox approach which suggested the application of a square root transformation to the total aphid counts. Tables \ref{tab:BRCM1}--\ref{tab:BRCNM_Farmers1} report the the results of nested model comparisons for the models BRCM, BRCNM (King's case) and BRCNM (Farmer's case) respectively, including two orders for fitting the direct treatment and network effects for each of the network effect specifications. This is because from a design perspective having the network effects first ties up with the optimality criterion used in this paper (i.e. estimate direct effects in the presence of nuisance network effects), but from a modelling perspective network effects are added second in order to understand what is their added contribution. Despite the complex blocking structure (for removing spatial variation) there is evidence that the network effects are important. Thus we can observe that the background variation is partly explained by allowing for the network effects, and this provides a justification for designing the experiment accounting for the network effects as well as the complex blocking structure. We can discern that the network effects are significant for the King's case but not for the Farmer's case suggesting that the assumption that some of the farm management activities might have influenced the responses is not likely. Recall that the King's case is obtained based on the weighted spatial structure capturing the neighbour effects from adjacent plots, while the farmer's case is directional and unweighted capturing the farmer activities. This might be because there is more association with the blocking structure for the farmer's case. However, if we assume that the best design (RCNBD) is much better than the design that was actually used, then the design that was used may not be powerful enough to detect these network effects even if they exist. The analysis of the motivating agricultural experiment provides evidence that the BRCNM captures important sources of interference by accounting for the network effects indicating the need to design accounting for these effects.

\clearpage

\bibliographystyle{asa}
\bibliography{bibliography} 

\end{document}